# From Range Loss to Recovery - Cold Weather Challenges and Design Strategies for Commercial Electric Vehicle Fleets

Soham Ghosh
Black & Veatch, Overland Park, Kansas - USA
sghosh27@ieee.org
ORCiD: 0000-0002-6151-8183
*Senior Member, IEEE*

Arpit Bohra
University of Texas at Austin, USA
arpitbohra@utexas.edu
ORCiD: 0000-0003-4793-863X
*Member, IEEE*

Karthik Saikumar
Arizona State University, USA
ksaikuma@asu.edu
ORCiD: 0009-0000-3255-1964
*Member IEEE*

*Abstract*—The North American commercial electric vehicle (EV) sector is undergoing rapid expansion, with unit sales rising from 21,120 in 2022 to 36,491 in 2023—a 73% increase, according to the International Energy Agency. However, this accelerating adoption brings emerging technical challenges. One critical concern is the impact of low to extreme winter temperatures (25°F to -25°F) on EV performance, including reduced energy efficiency and extended charging times. This paper presents a systematic analysis of commercial EV performance degradation under cold weather conditions and its broader implications on grid operations. Monte Carlo simulations, applied using real-world fleet parameters, indicate that approximately 200 MWh of additional daily energy demand may be required in the U.S. alone to offset efficiency losses during severe cold events. The resulting strain on an already stressed winter grid could exacerbate reliability risks. Moreover, increased harmonic distortion associated with cold weather charging behaviors has also been observed, raising concerns about power quality. To address these challenges, this study proposes two practical mitigation strategies: (1) a 'design-integrated safety' battery swapping station model operating in thermally controlled environments to significantly reduce charging downtime, and (2) a hybrid architecture combining roadside fast charging with depot-based deep charging to support continuous fleet utilization without compromising range. Together, these interventions provide a robust foundation for resilient commercial EV integration in cold climates, supporting fleet operators and utilities in managing seasonal performance variability.

## I. INTRODUCTION

Electric vehicle (EV) adoption has surged globally, driven by a combination of climate policy, technological advancement, and market incentives. China continues to lead this transition, contributing nearly 60% of global EV sales, with electric vehicles making up 22% of its domestic vehicle market. In Europe, Norway, Iceland, and Sweden are setting the pace, with EVs accounting for 80%, 41%, and 32% of new passenger vehicle sales, respectively [1]. In contrast, the United States, despite being the world's third-largest automotive market, lags behind [2] particularly in the commercial EV segment. Commercial EVs remain relatively expensive compared to internal combustion engine (ICE) counterparts, prompting federal and local governments to introduce tax credits, subsidies, and rebates. However, financial incentives alone have not sufficiently accelerated commercial EV deployment. A key barrier persists in the form of operational limitations, particularly related to vehicle range and reliability in cold climates. This manuscript focuses on the niche area of commercial EVs and the adoption barriers it faces as we try to extend their operations in cold weather conditions.

### A. Research motivation

The push toward electrification is reinforced by zero-emission targets and net-zero pledges from countries such as Canada, the UK, and Nordic nations, all of which are accelerating efforts to decarbonize commercial transportation. Despite these ambitions, range anxiety remains a major concern, especially for operators managing fleets in rural, regional, or extreme weather environments. While the non-commercial EV space has benefitted from substantial research into range extension technologies, battery advancements, and widespread charging infrastructure, commercial EVs, especially in North America, have not experienced similar momentum. This disparity is most evident under cold and extreme cold conditions, where commercial EVs suffer from reduced range, prolonged charging times, and diminished battery efficiency. Compounding the issue, commercial EVs operate under markedly different duty cycles and load conditions compared to passenger vehicles, requiring distinct cold weather hardening strategies. Addressing this gap is critical for ensuring the viability, resilience, and scalability of zero-emission commercial transportation systems.

### B. Manuscript contributions

This study advances the understanding in the resilient commercial EV charging/ infrastructure space particularly under extreme cold weather conditions, offering both empirical insights and deployable solutions. The key contributions are as follows:

1) *Cold weather SoC characterization:* Experimental results under ambient temperatures of −25°F reveal excessive charging times and a levelling of the battery state of charge (SoC) at approximately 80%, highlighting thermal limitations in subzero environments.

2) *Monte Carlo simulation of grid impacts:* Simulations quantifying the additional energy (in MWh) required to overcome cold weather charging inefficiencies with aggregation of commercial EVs in the US, thereby demonstrating potential grid reliability challenges under scenarios of high commercial EV fleet penetration.

3) *First-of-a-Kind (FOAK) battery swapping station design:* A novel and energy efficient swappable battery station architecture is introduced, incorporating 'design-integrated safety' thermal management and fire containment features to mitigate risks from battery thermal runaway events.

4) *Hybrid charging strategy for transit electrification:* A commercially viable hybrid charging model, combining strategically placed fast-charging stops and slower overnight depot charging. The hybrid charging model is demonstrated through a conceptual real-world case study of EV transit deployment at Chicago O'Hare (ORD) airport.

The manuscript is structured as follows: Section II examines the impact of extreme cold weather on commercial EV fleet performance and grid reliability, emphasizing increased energy consumption and elevated current total harmonic distortion levels. Section III presents the advantages of battery swapping stations and introduces a proof-of-concept architecture to address thermal runaway risks. Section IV proposes a hybrid charging strategy, combining fast charging at strategic stops with slower overnight depot charging, as a practical solution to mitigate cold weather-induced range loss. Finally, Section V summarizes the key findings, outlines current limitations, and identifies directions for future research.

## II. Impact of Cold Weather on EV fleet and Grid Performance

### A. Commercial EV adoption – categorization and trends

The adoption of commercial electric vehicles (EVs) is accelerating globally as fleet operators seek to reduce operational emissions and comply with tightening regulatory standards. Commercial EVs can broadly be categorized into four main segments, see Fig. 1: (1) urban light-duty delivery vehicles, which are ideal for last-mile logistics in dense city environments; (2) regional medium-duty transport trucks used for intercity freight and services; (3) urban electric buses serving municipal and airport transit needs; and (4) long-haul electric semi-trucks designed for interstate commerce. Each category presents unique operational demands and infrastructure challenges, particularly related to range, payload, and charging logistics. The shift toward electrification in these sectors is being driven not only by falling battery costs and improvements in drivetrain efficiency, but also by rising pressure to decarbonize transportation systems - one of the largest sources of greenhouse gas emissions globally.

According to data from the International Energy Agency's Global EV Data Explorer, China has already achieved substantial electrification in its commercial fleet, with 64% of its buses being electric as of 2024, a figure projected to exceed 70% by 2030. In contrast, the global average for electric bus adoption is expected to reach only 17% by that time. Similarly, China is projected to reach a 50% adoption rate for electric light- and medium-duty trucks and vans by 2030, while the global average for this segment is forecasted at just 25%. These disparities underscore China's outsized role in driving the pace and scale of commercial EV deployment worldwide.

| Commercial EV category | Typical range (miles) | Typical battery capacity (kwh) | Primary use case | Payload sensitivity |
|---|---|---|---|---|
| 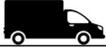 Urban light-duty delivery vehicles | 80-150 | 40-80 | Last-mile deliveries in urban areas | Low to moderate |
| 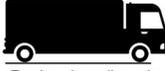 Regional medium duty transport truck | 150-250 | 150-300 | Intercity goods transport and logistics | Moderate |
| 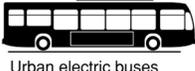 Urban electric buses | 120-250 | 250-400 | City transit, airport shuttles | Moderate to high |
| 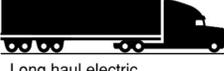 Long haul electric semi-trucks | 300-500 | 500-1000 | Interstate freight and logistics | High |

Fig. 1. Technical and use case overview of major commercial EV categories.

China, has currently established itself as the global leader in commercial EV deployment and projected to maintain the leading role (Fig. 2) is setting an influential precedent through its aggressive policy mandates and technological advancements. The country is significantly tightening emissions regulations for heavy- and medium-duty trucks, incentivizing a rapid transition to zero-emission alternatives. These developments are likely to shape the regulatory landscape and market trajectory in Europe and North America, where similar environmental goals are being prioritized. China's advancements in battery technologies, scalable charging infrastructure, and manufacturing efficiency offer a critical roadmap for other regions aiming to accelerate commercial EV adoption while navigating the complexities of vehicle deployment, grid readiness, and cross-sector coordination.

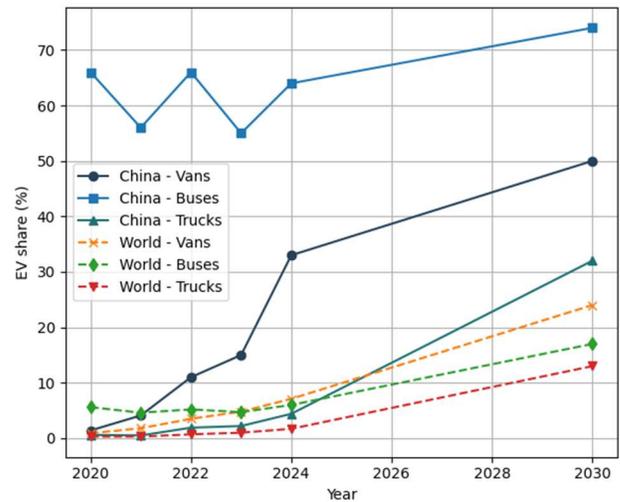

Fig. 2. Commercial EV sales share projection through 2030 by vehicle types (China vs world) [3].

## B. Impact of cold weather on EV fleet range and charging time

Cold weather presents a significant challenge to the performance and charging behavior of lithium-ion (Li-ion) batteries used in electric vehicles (EVs). As ambient temperatures fall, the electrolyte inside the battery becomes increasingly viscous, impairing lithium-ion mobility between the anode and cathode. This reduced ion transport leads to a marked increase in internal resistance, which diminishes both the battery's power output and charging efficiency. Notably, Chacko and Chung's experimental findings [4] reveal that a drop in temperature from 113°F to 10°F can elevate a battery's internal resistance by as much as five times its nominal value. This increased resistance not only slows down the rate at which energy can be stored in the battery, but also necessitates more conservative charging protocols to prevent degradation and ensure safety.

At extremely low temperatures (-10°F and below), such as those encountered during winter storms or arctic climates, EV batteries often struggle to reach a full 100% state of charge. This is not solely due to the electrochemical limitations from increased resistance but also from the compounding demands of the vehicle's auxiliary systems. Specifically, the battery thermal management system draws a significant amount of power to maintain cell temperatures within operational limits, while interior cabin heating further reduces net available energy. Under experimental setup, these parasitic loads combined with reduced electrochemical efficiency, resulted in longer charging durations and lower achievable charge levels, frequently capping the battery's SOC at around 80–85% under severe cold conditions, as one may observe in Fig. 3. Understanding these dynamics is crucial for infrastructure planning and grid integration strategies, particularly in regions experiencing cold climates with growing commercial EV adoption.

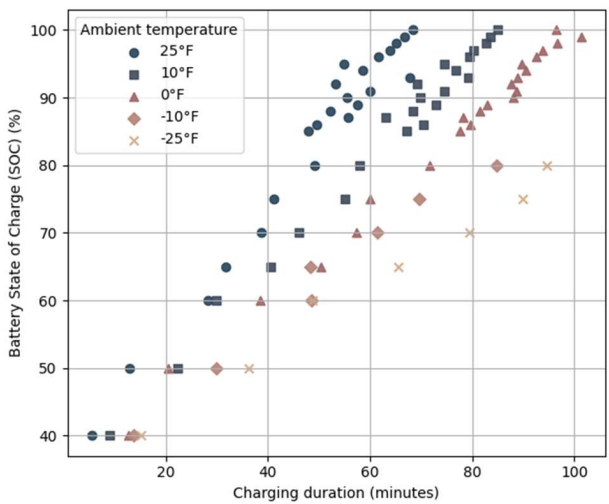

Fig. 3. Battery SOC versus charging duration under different experimental ambient temperatures. Observe battery non-performance at -10°F, -25°F.

To address the limitations posed by cold weather on EV charging performance, it is important to consider the practical differences between non-commercial and commercial EV applications. While traditional modifications, such as using heated seats and steering wheels to reduce cabin heating loads [5], may help conserve energy in personal EVs, these measures are often impractical for commercial fleets, particularly passenger shuttle buses, where maintaining interior comfort is essential. For these applications, especially at temperatures above 10°F [6], heat pump systems offer a more effective alternative to resistive heating to maintain the effectiveness of the battery. Heat pumps not only reduce auxiliary power draw but also help protect batteries from thermal degradation and improve charging efficiency, a strategy beneficial and implementable across both commercial and non-commercial EV platforms.

## C. Impacts of cold weather on grid power consumption

A Monte Carlo simulation was conducted to estimate the increased power demand from commercial electric vehicles (EVs) in the United States under cold weather conditions. The simulation assumed a national fleet of 28,200 commercial EVs with a total available battery capacity of 5.9 GWh, based on 2024 data obtained from the International Energy Agency's *Global EV Data Explorer* [3] repository. Each vehicle was assumed to travel an average of 100 miles per day, with baseline energy consumption calculated using an average efficiency of 2.5 miles per kilowatt-hour (mi/kWh), resulting in a daily demand of approximately 40 kWh per vehicle. Cold weather effects were modeled as stochastic efficiency losses, ranging from 0–15% for average cold conditions (25.9°F) and 10–35% for severe cold (-10°F). To reflect operational constraints, the simulation imposed a minimum state-of-charge (SoC) of 40%, capping daily usable battery energy at 60% of capacity. Over 10,000 simulation iterations, the mean daily fleet energy demand increased from approximately 1.22 GWh (mean) under average cold conditions to 1.47 GWh (mean) during severe cold, see Fig. 4, representing a ~20% rise in energy consumption. These results highlight the significant impact of subfreezing temperatures on EV fleet efficiency and resulting grid load.

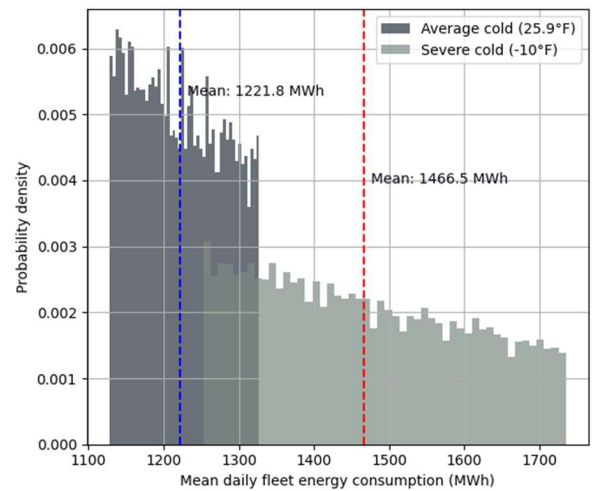

Fig. 4. Distribution of mean daily fleet energy consumption for commercial EVs in cold conditions.

This surge in electricity consumption due to efficiency losses poses a critical challenge during periods when the grid is already under stress, where prolonged subfreezing temperatures put strains on both supply and infrastructure. Without appropriate planning for peak seasonal loads, high commercial EV penetration could exacerbate reliability risks during cold snaps, similar to the winter energy crisis Texas experienced in 2021 during the winter storm Uri [7, 8].

### D. Impacts on grid from higher current THD from cold weather charging

The impact of electric vehicle (EV) charging on power distribution networks is a well-explored domain, with a substantial body of literature addressing load dynamics, peak demand, and infrastructure implications. However, a critical gap remains in understanding the influence of ambient temperature on total harmonic distortion (THD) - particularly current THD - during EV charging. Except for a few isolated studies, notably [9, 10], the relationship between temperature variation and harmonic behavior remains largely unexplored. Notably, there is a complete absence of data analyzing how cold climates affect THD in chargers rated at 50, 75, and 100 kW, power levels commonly used in commercial fleet applications. To address this shortfall, this study conducted controlled measurements of current THD across four commercially available DC fast chargers (50, 75, 100, and 150 kW), operated under climate- and humidity-controlled conditions with a fixed minimum state-of-charge (SoC) threshold of 40%. Three ambient temperature setpoints, 25°F, 0°F, and -25°F, were evaluated. Results, in Fig. 5, indicate a consistent and measurable increase in current THD as ambient temperature decreases, suggesting that harmonic emissions may become more pronounced under cold-weather charging conditions. An important observation from the experiment was that though the vendor datasheets claim wide operating ranges from, typically ranging from -31°F to 122°F, they provide limited assurance that harmonic performance shall remain within acceptable thresholds per IEEE 519 or IEC 61000-3-12 standards at the lower temperature end of this spectrum. Higher $THD_I$ underscores two key implication:

1) The need for expanded testing protocols and updated compliance documentation to ensure power quality is not compromised as commercial EV adoption expands into colder climates.

2) The need for potentially testing cold weather harmonic signatures, at the main service entrance of charging station depots, especially for the depots that operate in cold weather conditions with outdoor charging stations.

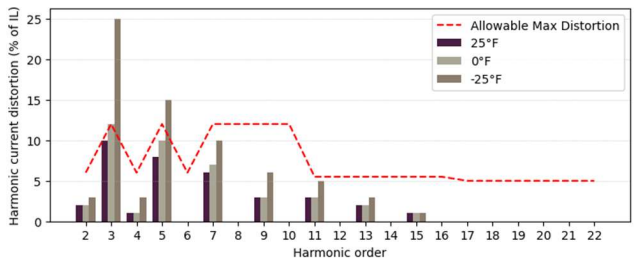

Fig. 5. Variation of $THD_I$ as a function of ambient temperature for non-compliant commercial charger(s), showcasing certain non-compliance with IEEE 519 THD limits at lower operating temperatures ($100<I_{sc}/I_L<1000$).

Now that the mechanics underlying the impacts of cold weather on the performance of EV fleet and the distribution power grid is understood, the following two sections present engineering solutions to recover the range loss for these fleet vehicles.

## III. RECOVERING RANGE LOSS THROUGH BATTERY SWAPPING STATION

Battery swapping has seen limited uptake in the private EV market due to issues related to battery ownership and the absence of standardized battery designs across vehicle manufacturers. For example, Better Place, which attempted to scale battery-swapping for private EVs, sold only about 750 cars in Israel before accumulating over $500 million in losses, highlighting the economic and logistical hurdles in that segment. However, in commercial fleet operations, where vehicles are centrally managed and often built on uniform platforms, these challenges are significantly reduced, making battery swapping a more practical and effective solution.

### A. Benefits of battery swapping stations for commercial EV fleets

Battery swapping offers several operational and economic advantages for commercial EV fleets over conventional plug-in or pantograph-based charging approaches:

- *Rapid turnaround time:* Battery swapping takes only a few minutes, typically under 5 minutes, compared to 30–90 minutes required for DC fast charging or several hours for conventional Level 2 charging, significantly reducing vehicle downtime and improving fleet utilization.

- *Enhanced battery longevity:* Since batteries in a swapping station can be charged gradually under thermally controlled conditions, their state of health (SoH) is better preserved. For instance, charging at slower rates (C/2 or lower) and maintaining temperatures between 20–25°C can extend battery cycle life by over 30% compared to uncontrolled fast charging in harsh ambient conditions.

- *Centralized monitoring and maintenance:* A battery swapping station consolidates operational complexity into a single facility, making it easier to manage safety, maintenance, and diagnostics compared to deploying and maintaining multiple dispersed fast-charging sites.

- *Time-of-use optimization:* Charging schedules within a swapping station can be aligned with electricity price signals, enabling batteries to be charged during off-peak hours when electricity costs are lower—reducing overall energy expenses for fleet operators.

As an use case, in a commercial battery swapping station capable of concurrently charging 36 batteries (each 400 kWh, as typical for electric bus fleets), diverting just 25% of the fleet (9 batteries or 3,600 kWh/day) to off-peak Time-of-Use (TOU) schedules can yield significant annual cost savings. With Illinois' average commercial electricity rate at approximately 13.13 ¢/kWh, and assuming a TOU discount of 3 ¢/kWh (i.e., off-peak at ~10¢/kWh versus average 13.13¢), the daily cost reduction is about $108, translating to around $39,400 in annual savings. Factoring in higher TOU spreads in specific contracts, practical savings could realistically range between $40,000 and $60,000 per year. These findings underline the economic benefits of integrating dynamic load management and TOU optimization into high-capacity commercial EV charging operations based on battery swapping.

- *Grid integration and revenue opportunities:* Large-scale battery swapping stations can actively participate in utility-led demand response programs or function as virtual power plants (VPPs). By aggregating idle batteries as distributed energy resources (DERs), these stations can supply ancillary services such as frequency regulation or peak shaving, providing grid stability while generating additional revenue streams for operators.

Now that the merits of battery swapping in the commercial EV space are established - from minimizing vehicle downtime to preserving battery health and enabling grid integration - the focus shifts to the critical aspect of ensuring operational safety. In the following section, a 'design-integrated safety' architecture is presented, addressing thermal management, fault isolation, and containment measures to mitigate the risks associated with battery thermal runaway events.

*B. Battery swapping station - 'design-integrated safety'*

Battery swapping stations operate by replacing a depleted battery with a fully charged one, while the drained battery is stored and recharged in a dedicated rack system. These rack arrangements typically house and charge dozens of lithium-ion batteries simultaneously, presenting significant safety challenges due to the risk of thermal runaway - an issue also observed in grid-scale battery energy storage systems (BESS). At the core of lithium-ion battery chemistry lie four key components: the positive and negative electrodes, a liquid electrolyte, and a microporous separator. The liquid electrolyte, commonly composed of flammable linear and cyclic carbonates, lithium salts, and additives, is particularly vulnerable under thermal or electrical stress. With flash points below 0°F [11], linear carbonates are highly flammable and represent a primary trigger for thermal runaway, especially during high-cycle operation without sufficient thermal management [12, 13].

While alternative electrolytes, such as ionic liquids, fluorinated solvents [14], flame retardants [15], and solid electrolytes comprising of polymer and organic or inorganic salts have been explored [16, 17], widespread adoption is hindered by trade-offs in cost, electrochemical performance, and scalability. Consequently, conventional carbonate-based electrolytes remain dominant in commercial applications. Once initiated, thermal runaway becomes a self-sustaining reaction that cannot be halted until all reactive components are consumed.

To address this critical safety risk, the proposed 'design-integrated safety' architecture, shown in Fig. 6, recommends housing battery racks within ISO containers equipped with top-mounted deflagration panels, in full compliance with NFPA 68 standards for deflagration venting. This containment approach is especially vital when battery swapping infrastructure is collocated with critical assets such as airports, enabling localized fault isolation and minimizing the potential for catastrophic damage. In addition to the safety aspect, the containerized charging solution proposed in this manuscript provides the added benefit of independent temperature control of these batteries from that of the ambient temperate of the facility, which would allow the facility temperate to be few degrees lower than that of the batteries, if needed, resulting in potential cost savings.

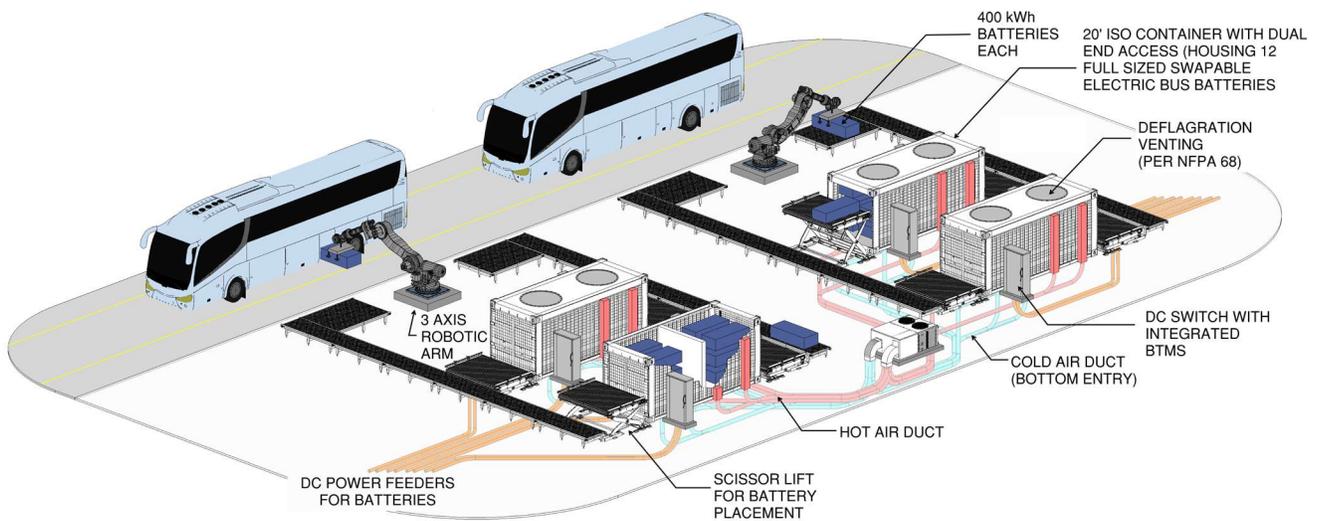

Fig. 6. Commercial EV battery swapping station: illustration comparing rack mounted battery charging versus contained based design with deflagration panel ('design-integrated safety').

## IV. Recovering Range Loss Through Hybrid Fast Charge Stops Paired with Depot Deep Charging

While battery swapping stations offer operational efficiencies such as reduced vehicle downtime, their deployment in commercial electric vehicle (EV) fleets, particularly electric buses, presents some logistical and infrastructural challenges. Accessing battery swapping infrastructure might require detours from fixed transit routes, which can disrupt service schedules and reduce route efficiency. This becomes especially problematic in high-demand, infrastructure-constrained environments like airport transit systems, where land use is tightly regulated and flexibility for retrofitting depot facilities to accommodate battery swapping infrastructure is limited.

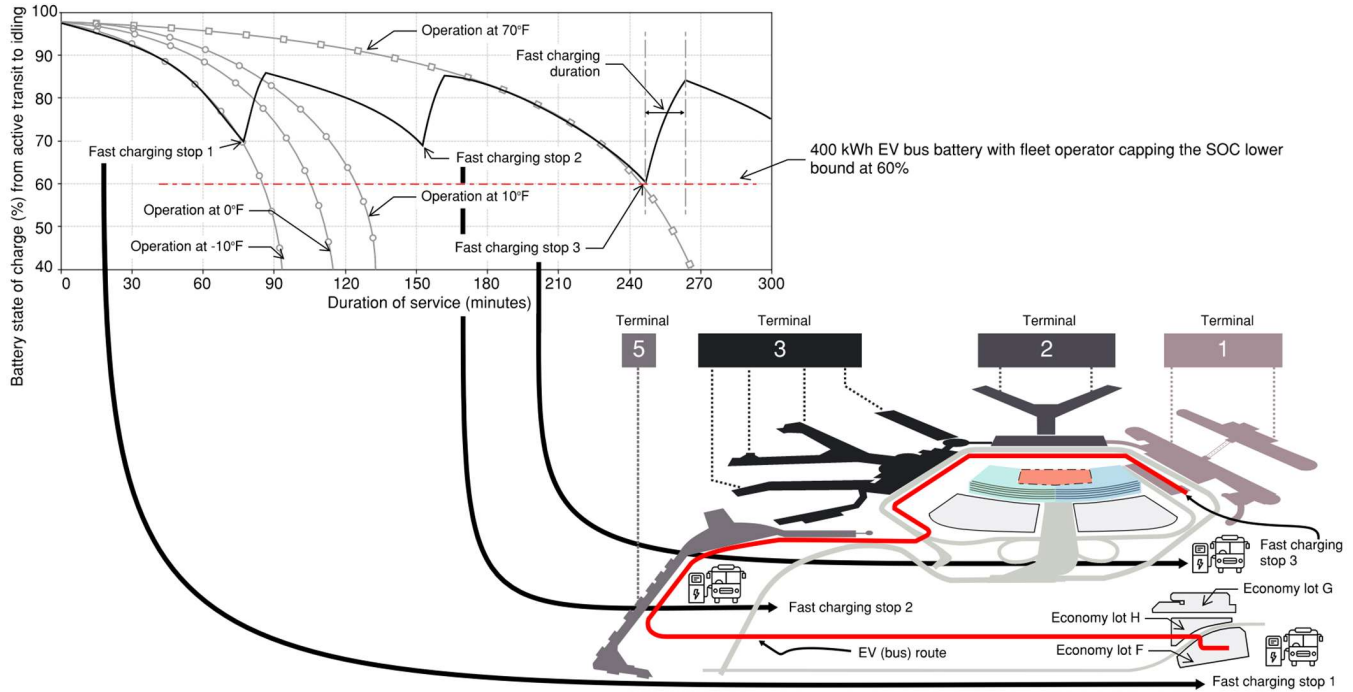

Fig. 7. Test bed similation of the conceptual Chicago O'Hare airport's EV bus transit with fast DC charge stops paired with overnight depot deep charging (depot charging route not shown in this illustration for simplicity).

In such scenarios, transit operators face difficult trade-offs to maintain service reliability. Two solutions primarily exists in this scenario:

1) One approach involves maintaining a parallel contingency fleet powered by compressed natural gas (CNG), which adds significant capital and operational costs, while also undermining emissions reduction goals.

2) Alternatively, agencies may deploy standby EV buses to compensate for extended charging durations, but this too imposes additional fleet size, storage, and maintenance burden - factors that collectively inhibit the scalability of EV adoption in public transportation networks.

An even more critical, long-term implication arises from the battery degradation dynamics associated with cold weather operations and frequent deep discharges. Repeated exposure to suboptimal temperatures combined with deep depth-of-discharge (DoD) events accelerates battery aging, reducing the state of health (SoH) significantly over time. Empirical study [18] suggest that operation at −10°F can reduce lithium-ion battery capacity retention by up to 20% over 1,000 cycles, compared to ambient conditions. This accelerated degradation imposes additional maintenance cycles and shortens the effective service life of EV batteries, further straining budgets and undermining operational stability.

A viable mitigation strategy under such constraints involves deploying DC fast charging stations at selected strategic points along a vehicle's route. This approach enables partial opportunity charging during layovers or low-occupancy periods, ensuring that the battery's state of charge (SoC) remains above critical thresholds, thereby minimizing deep discharges and extending battery lifespan. Overnight depot-based slow charging can then be used for full charge cycles at a lower cost and without additional demand charges on the electric grid.

To illustrate this approach, a conceptual electric bus fleet strategy at Chicago's O'Hare International Airport (ORD) serves as a representative case. The Chicago Transit Authority (CTA), as outlined in its "Charging Forward" strategic roadmap [19], aims to transition to a fully electric bus fleet by 2040. This involves a multi-pronged investment in vehicle procurement, facility upgrades, and a diversified charging infrastructure. Simulations conducted at an ambient temperature of 70°F for CTA's airport-serving bus routes show in Fig. 7, an average continuous operational duration of approximately 270 minutes before the battery depletes to a lower limit of 40% SoC. However, in colder conditions (−10°F), this runtime drops drastically to about 90 minutes, highlighting the severe impact of cold weather on EV performance.

By integrating three strategically placed 50 kW Veefil Tritium DC fast chargers (charger model selected for charging discharging simulation only), at the economy F parking lot, Terminal 5, and Terminal 3 stops, transit planners can introduce short-duration (10–15 minute) high-power charging sessions during scheduled stops. These allow the battery SoC to be replenished to approximately 85%, while avoiding critical depletion levels below 40%. This dual-layered charging strategy, i.e. on-route DC fast charging combined with slower overnight depot charging – is aimed to support continuous operations while preserving battery health, offering a replicable model for other transit agencies facing similar constraints.

The proposed architecture in this section can be adopted to other major airports and in essence consists of strategically placed 50 kW fast chargers to enable continuous electric bus operation without depot returns or natural gas backups, even during extreme cold.

## V. Summary, Challenges, and Future Work

This paper provides a comprehensive review of current and projected trends in commercial electric vehicle (EV) adoption, with a particular emphasis on regional disparities and climatic challenges. The analysis underscores how China's accelerated deployment of commercial EVs, particularly buses and medium-duty trucks, positions it with a strategic advantage. With over 64% of its bus fleet already electrified as of 2024 and projections exceeding 70% by 2030, China sets a global benchmark for fleet-scale decarbonization. These insights serve as a comparative backdrop to highlight the emerging constraints and opportunities for commercial EV adoption in North America and Europe.

A critical barrier identified for these regions is the operational and grid-level vulnerability introduced by extreme cold weather. Through experimental testing and simulation-based modeling, this work demonstrates how sub-zero ambient temperatures degrade state of charge performance, reduce usable range, and exacerbate grid stress due to increased charging demands. To address these challenges, two actionable infrastructure solutions have been proposed: (1) a 'design-integrated safety' battery swapping station architecture, equipped with deflagration venting for risk containment in dense urban environments, and (2) a hybridized charging strategy that combines fast charging at key transit stops with slower overnight depot charging. These approaches are especially suited for urban buses and light- to medium-duty commercial fleets operating within metropolitan and regional corridors.

The applicability of the strategies discussed in the manuscript to long-haul electric semi-trucks remains limited due to their distinct energy requirements and duty cycles. This represents a key limitation of the current work and an important direction for future research. Additional investigations are also warranted to enhance the resilience of commercial EV chargers under extreme temperatures, ensuring both performance reliability and compliance with power quality standards. Further, integrating high-efficiency thermal management systems, such as heat pumps, into electric buses could play a pivotal role in preserving range and operational efficiency in cold climates, marking another avenue for strategic innovation in commercial EV fleet development.

## Declaration of Interest and Acknowledgements

The views, thoughts, opinions, conclusions made in this material are solely those of the authors and don't necessarily reflect the view of the authors' employer, organization, committee or other group or individual. The findings in this paper were not sponsored by any agency or institution.

The authors would like to thank Chip Blackwell for his expertise in helping us with certain visualizations for the manuscript.